\documentclass{sig-alternate-05-2015}
\usepackage{graphicx} % for EPS use the graphics package instead
\usepackage{balance}  % for useful for balancing the last columns
\usepackage[show]{chato-notes}
\begin{document}

%\title{A Hybrid Approach for Automatic Classification of SMS Messages}

%\conferenceinfo{Submitted for review}{Jan 2016}
%\permission{Submitted for confidential review.}
%\copyrightetc{January 2016}
\CopyrightYear{2016}
\setcopyright{acmcopyright}
\conferenceinfo{DH '16,}{April 11-13, 2016, Montr\'{e}al, QC, Canada}
\isbn{978-1-4503-4224-7/16/04}\acmPrice{\$15.00}
\doi{http://dx.doi.org/10.1145/2896338.2896364}

\clubpenalty=10000 
\widowpenalty = 10000

\title{Enabling Digital Health by Automatic Classification of Short Messages}

\newcommand*\samethanks[1][\value{footnote}]{\footnotemark[#1]}
\numberofauthors{5}
% Notice how author names are alternately typesetted to appear ordered
% in 2-column format; i.e., the first 4 autors on the first column and
% the other 4 auhors on the second column. Actually, it's up to you to
% strictly adhere to this author notation.
\author{%
  \alignauthor{%
Muhammad Imran\\
    \affaddr{Qatar Computing Research Institute, HBKU} \\
    \affaddr{Doha, Qatar} \\
    \email{mimran@qf.org.qa} }
\alignauthor{%
Patrick Meier\thanks{Work done while the author was at QCRI.}\\
    \affaddr{World Bank}\\
    \affaddr{Washington DC, USA}\\
    \email{patrick@irevolutions.org} }
\alignauthor{%
Carlos Castillo\samethanks\\
    \affaddr{Eurecat}\\
       \affaddr{Barcelona, Spain}\\
       \email{chato@acm.org}}
\and
\alignauthor{%
Andre Lesa\\ 
    \affaddr{UNICEF}\\
    \affaddr{Zambia}\\
    \email{andre.lesa@live.com}}
\alignauthor{%
Manuel Garcia Herranz\\
    \affaddr{UNICEF}\\
    \affaddr{Zambia}\\
    \email{mgarciaherranz@unicef.org} } }

% Paper metadata (use plain text, for PDF inclusion and later
% re-using, if desired)
%\def\plaintitle{SIGCHI Extended Abstracts Sample File: Note Initial
%  Caps} \def\plainauthor{First Author, Second Author, Third Author,
%  Fourth Author, Fifth Author, Sixth Author}
%\def\plainkeywords{Authors' choice; of terms; separated; by
%  semicolons; include commas, within terms only; required.}
%\def\plaingeneralterms{Documentation, Standardization}

%% Set up our PDF with metadata
%\hypersetup{%
%  pdftitle={\plaintitle}, pdfauthor={\plainauthor},
%  pdfkeywords={\plainkeywords}, }

% \reversemarginpar%

%\begin{document}

%\setcopyright{acmcopyright}

\maketitle

% Uncomment to disable hyphenation (not recommended)
% https://twitter.com/anjirokhan/status/546046683331973120
%\RaggedRight{} 

% Do not change the page size or page settings.
\begin{abstract}
In response to the growing HIV/AIDS and other health-related issues, UNICEF through their U-Report platform receives thousands of messages (SMS) every day to provide prevention strategies, health case advice, and counseling support to vulnerable population. Due to a rapid increase in U-Report usage (up to 300\% in last 3 years), plus approximately 1,000 new registrations each day, the volume of messages has thus continued to increase, which made it impossible for the team at UNICEF to process them in a timely manner. In this paper, we present a platform designed to perform automatic classification of short messages (SMS) in real-time to help UNICEF categorize and prioritize health-related messages as they arrive. We employ a hybrid approach, which combines human and machine intelligence that seeks to resolve the information overload issue by introducing processing of large-scale data at high-speed while maintaining a high classification accuracy. The system has recently been tested in conjunction with UNICEF in Zambia to classify short messages received via the U-Report platform on various health related issues. The system is designed to enable UNICEF make sense of a large volume of short messages in a timely manner. In terms of evaluation, we report design choices, challenges, and performance of the system observed during the deployment to validate its effectiveness.
\end{abstract}

%\keywords{text classification, hybrid system, stream classification, supervised machine learning}

%\category{H.5.m}{Information interfaces and presentation (e.g.,
%  HCI)}{Miscellaneous}\category{See}{\url{http://acm.org/about/class/1998/}}{for
%  full list of ACM classifiers. This section is required.}
%
% The code below should be generated by the tool at
% http://dl.acm.org/ccs.cfm
% Please copy and paste the code instead of the example below. 
%

%\begin{CCSXML}
%<ccs2012>
%<concept>
%<concept_id>10002951.10003227.10003245</concept_id>
%<concept_desc>Information systems~Mobile information processing systems</concept_desc>
%<concept_significance>300</concept_significance>
%</concept>
%</ccs2012>
%\end{CCSXML}
%
%\ccsdesc[300]{Information systems~Mobile information processing systems}

%
% End generated code
%

%
%  Use this command to print the description
%
%\printccsdesc

% We no longer use \terms command
%\terms{Theory}


% !TEX root = paper.tex

\section{Introduction}
In response to the growing HIV/AIDS and other health-related issues, UNICEF, under the leadership of the Zambian government, piloted the U-Report project\footnote{http://zambiaureport.com/} (initiated in Uganda~\cite{melville2013amplifying}) over the past year, leveraging mobile phone technology to distribute confidential prevention strategies, health care advice, and counseling support to vulnerable populations in hard-to-reach areas. U-Report is a free and open-source SMS platform used by UNICEF to engage on a wide variety of issues across different countries. UNICEF Zambia receives a large number of messages daily related to different kinds of health issues. 

Through the U-Report platform, UNICEF provides two-way correspondence between at-risk youth with access to a mobile device and health counselors trained in HIV/AIDS prevention and care strategies as well as conducted tailored demand creation polls and campaigns. Currently, around 90,000 U-Reporters have joined the service and they continuously send questions concerning public health, and specifically questions related to HIV/AIDS, to health counselors via SMS. UNICEF experts in turn responded with health advice and support by direct SMS. With the rapid growth of mobile technology, the use of the U-Report service has increased almost 300\% in just 3 years, plus approximately 1,000 new U-Reporters joining everyday. The volume and velocity of text messages received via U-Report has thus continued to increase, which has made it more difficult for the team at UNICEF Zambia to make sense of these messages in a timely manner. 

\begin{figure*}[t]
\centering
  \includegraphics[width=0.85\textwidth]{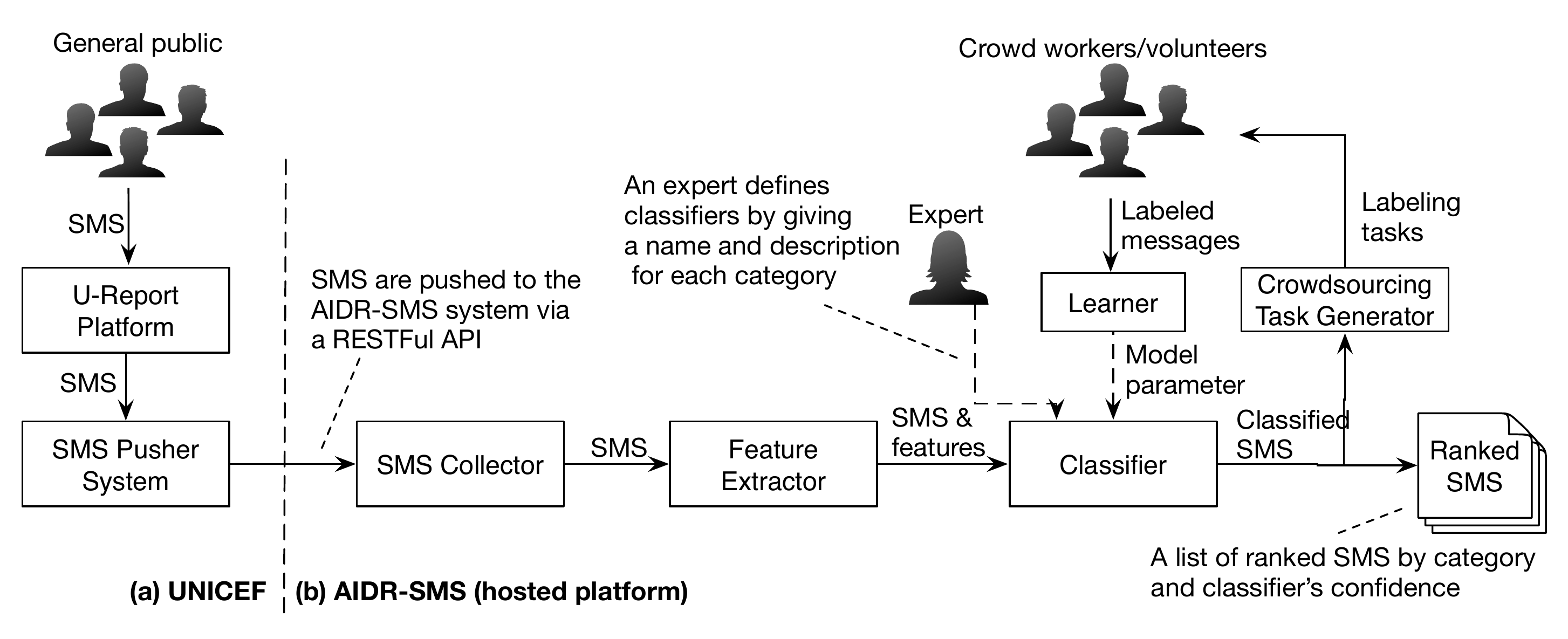}
  \caption{High-level architecture of the AIDR-SMS platform}~\label{fig:system}
\end{figure*}

In this paper, we present an open-source system, which combines human and machine intelligence to enable the real-time classification of short messages (SMS) through supervised machine learning techniques. The automated system seeks to resolve the information overload issue by introducing real-time processing of large-scale data at high-speed while maintaining high classification accuracy. We have recently tested the system in conjunction with UNICEF in Zambia to classify short messages to various health-related issues received via the U-Report platform. The system helps UNICEF experts categorize incoming messages into health-related categories (described in next section), which then enables the experts to respond in a timely manner.

The rest of the paper is organized as follows. Next, we describe how UNICEF categorize SMS messages through manual approaches. Section~\ref{sec:sys}, presents the proposed system, along with design principles and its architecture. We provide pilot study and evaluation details in section~\ref{sec:study}. Related work is described in section~\ref{sec:relatedwork}. And, we conclude the paper in section~\ref{sec:conclusion} along with future work.

% !TEX root = paper.tex

\section{UNICEF SMS Categorization \\ Approach}\label{sec:u-report}
According to UNICEF, every hour, about three Zambian youth aged 15-24 years old get infected with HIV.
Due to lack of comprehensive HIV prevention information and being not properly quipped with HIV high impact prevention services such as condoms, male circumcision, young people are more vulnerable\footnote{http://www.unicefstories.org/2014/05/16/ureport-providing-counseling-services-on-hiv-and-stis/}. 
The U-Report deployment in Zambia provides confidential, free of charge, counseling services on HIV and STIs to vulnerable population. This open source SMS-based system is a vital doorway to information for young people who might not have access to an internet cafe or mobile data.

%\begin{figure}[htbp]
%  \includegraphics[width=1\columnwidth]{figures/registration_vs_converstation}
%  \caption{U-Report registrations vs. converstations initiated by the UNICEF team from Jan-2013 to Jan-2016 reported by UNICEF~\protect\footnotemark}~\label{fig:system}
%\end{figure}
%
%\footnotetext{http://zambiaureport.com/web/metrics/}

The U-Report system receives SMS sent by people through an SMS gateway, as depicted in Figure~\ref{fig:system}(a). An SMS consists of textual content (140 characters) and some meta-data (e.g. sender information). UNICEF experts analyze each message before assigning an appropriate category to it---a list of categories is given below. For example, a received SMS {\tt Where does HIV come frm? Does HIV act like other diseases} is categorized as ``Transmission", and {\tt Can an STI result in2 HIV if not treatd on time?} as ``Treatment" then assigned to an appropriate team for a response.

%\paragraph{UNICEF categories}
\subsection{UNICEF categories}
UNICEF uses the following list of categorize for SMS classification:

\begin{itemize}
\item {\bf Symptoms:} Messages that primarily describe a person's symptoms in addition to questions on how these symptoms may relate to known diseases such as HIV.

\item {\bf Definition:} Questions that are primarily requesting the definition of different Sexual Reproductive Health terms.

\item {\bf Male Circumcision:} If a message contains questions on where to go for MC, procedure, cost, healing after the procedure.

\item {\bf Testing HIV:} If a message contains questions primarily about testing for HIV/AIDS.

\item {\bf Treatment:} If a message contains questions primarily about available treatments, treatment procedure.

\item {\bf Pregnancy:} If a message contains questions around pregnancy, prevention of, signs of and how to handle it.

\item {\bf Transmission:} If a message contains questions on how HIV/AIDS is transmitted, carriers and protective measures that can be undertaken.

\item {\bf Prevention:} If a message has questions mainly around prevention of HIV/AIDS infection such as with condoms, effectiveness of different prevention methods.

\end{itemize}

\subsection{Domain-specific challenges}
With the rapid adoption of the U-Report system and its growing usage, manual classification of messages is no longer feasible. Moreover, other approaches such as keyword based matching, cannot guarantee the coverage of the returned results~\cite{melville2013amplifying}.

Typically, SMS messages are brief, informal, unstructured and often contain misspellings and grammatical mistakes. Often, abbreviations and acronyms are used to shorten the message to fit limited length restriction. Due to these issues, in general, short text classification is a challenging problem~\cite{sriram2010short}. In particular, classification of data streams carrying items consisting of short text brings challenges to machine learning techniques. For example, a machine learning model trained on past labeled data may produce undesirable results when underlying concepts change. For instance, a symptoms classifier (as listed in the above categories) may start producing errors when a new, unseen type of symptom appear. This requires re-training of the model with fresh labeled data to adapt to new changes in a category.

To address these issues, in the next section, we present a hybrid system for the automatic classification of short text messages arriving in a streaming way.

% !TEX root = paper.tex
\section{Proposed system}\label{sec:sys}

\begin{figure}[t!]
\centering
    \includegraphics[width=\columnwidth]{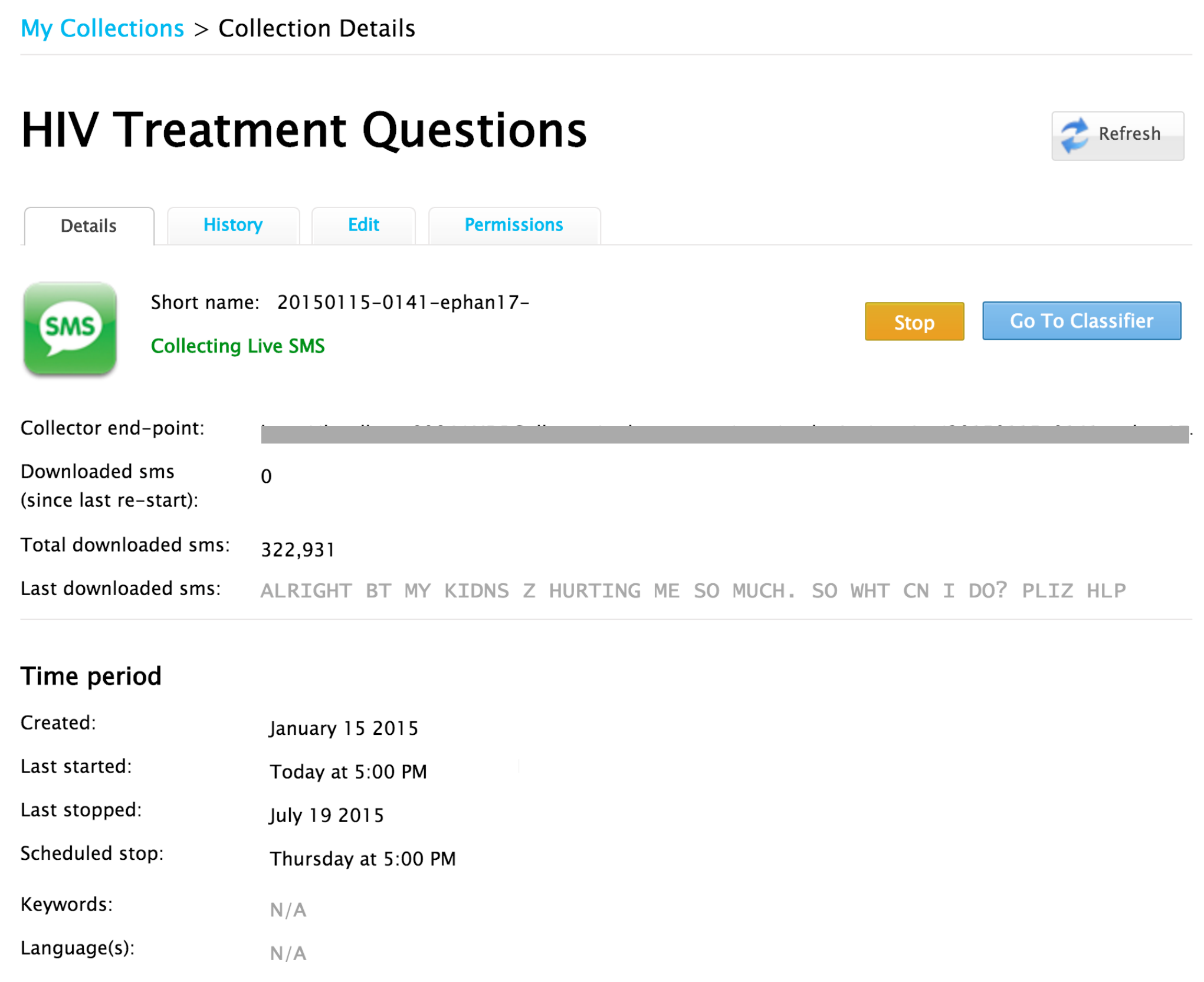}
    \caption{Collection interface showing various details of a SMS collection (end-point link is masked)\label{fig:collector}}
\end{figure}

\subsection{Design challenges}
Humans alone cannot be employed for the classification of continuous data streams at high-speed, because of the filter failure issue~\cite{imran2014coordinating}. Real-time processing of data requirement suggests to use machine intelligence for an accurate and faster response. For this purpose, we address the issue through the lens of crowd computing (i.e. crowdsourcing, microtasking, etc.) and machine computing (i.e. natural language processing, machine learning). 

Crowdsourcing is a new computing paradigm that can be used to address problems that are difficult for machines~\cite{geiger2011crowdsourcing}. In this case, typically a task is assigned to a crowd of workers by splitting it into several sub-tasks. Whereas, machine intelligence refers to automated algorithms for data processing. Indeed, automated data processing algorithms are much more faster than systems based on humans, however, often for difficult tasks such as image recognition, real-time document syntax checking, machines alone do not produce precise results.

Therefore, our proposed system is designed to combine both human and machine intelligence to address the issues that machines or humans cannot address alone. 
    
Among other design requirements, we expect the system to be responsive. That is, the system should maintain low latency and high throughput. A low latency system takes less time to deliver output than a high latency system. And, throughput is the speed at which items are processed: a high-throughput system can process more items per unit of time than a low-throughput one. Finally, the system must maintain a high quality (i.e. high classification accuracy), which is why we employ human intelligence to help machine when required.

\begin{figure}[t!]
    \includegraphics[width=\columnwidth]{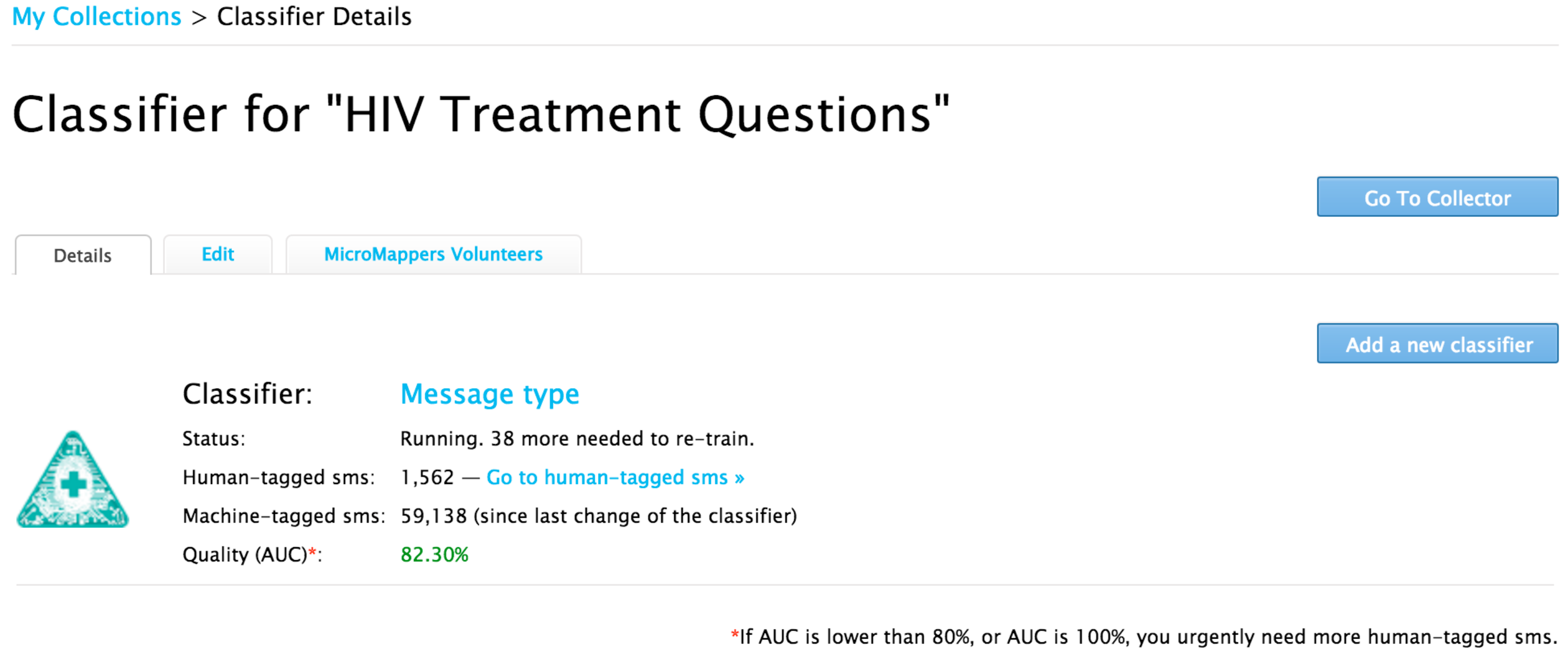}
    \caption{Classifier details interface showing machine classification accuracy in term of AUC and other details regarding human and machine tags \label{fig:classifier}}
\end{figure}

\subsection{System architecture}
To enable the automatic classification of SMS, we present AIDR-SMS\footnote{\url{http://aidr.qcri.org/}}, an SMS classification system developed as an extension of the baseline AIDR platform~\cite{imran2014aidr}. The system combines human and machine intelligence by bringing together characteristics from crowdsourcing and supervised stream classification systems. 

A high-level architecture of the system is shown in Figure~\ref{fig:system}. To process SMS received by the U-Report system, these messages should be directed to the AIDR-SMS system as soon as they arrive. For this purpose, we developed an SMS pusher system, which is deployed at the UNICEF server and responsible for sending messages to the AIDR-SMS system as soon as a new message is received by the U-Report platform. 

To initiate this communication between U-Report and AIDR-SMS, an expert performs the following steps through an online interface of the AIDR-SMS system:

\begin{itemize}

\item First, the expert defines an SMS collection in AIDR-SMS and gets a dedicated RESTFul API link of the collection. Multiple collections can be defined to handle multiple streams of messages. For example, this feature can be useful if there are multiple SMS short-codes setup by UNICEF. Figure~\ref{fig:collector} shows the interface of an SMS collection.

\item  Second, the expert creates custom classifiers. A classifier is a set of categories to which the messages will be classified by the system. Each category has a name and a description. For instance, in this case, all categories listed in section 2.1 are used create a classifier. Figure~\ref{fig:classifier} depicts the interface of a classifier created for SMS classification.
\end{itemize}

Once the above two steps are completed and the API link is provided to the SMS pusher system, the pusher is ready to start pushing messages to the AIDR-SMS classification system that it receives from the U-Report platform. This design enable AIDR-SMS to be used for processing several data streams in parallel.

As depicted in Figure~\ref{fig:system}, the \textit{SMS collector} module is responsible to receive messages from the SMS pusher. Also, it adds system-specific meta-data to each incoming message. Messages are then passed to the \textit{feature extractor} module, which generates content based feature vectors of the incoming messages. We use uni-grams and bi-grams as our features. The feature selector module is also responsible to select top 800 features. For this purpose, we use \textit{''Information gain"}, a famous feature selection method.

The \textit{classifier} module uses a supervised machine learning algorithm, namely \textit{Random Forest}~\cite{liaw2002classification} to train multi-class classifiers. As labeled examples are required for machine training, for this purpose, we use humans in the loop using the \textit{crowdsourcing task generator} module. The \textit{crowdsourcing task generator} module is responsible to select labeling tasks from the stream of messages to be labeled by human labelers. In our case, UNICEF experts perform all the labeling tasks. A labeling task, in this case, consists of an SMS message and a list of the categories (i.e. a category from the list of UNICEF categories shown in section 2.1) along with their description. The labeler reads the message and chooses one of the categories that he/she thinks most suitable for the message. A labeling task is finalized (i.e. a category is assigned to the message), if two out of three human labelers agreed on a label. 

\begin{figure}
    \includegraphics[width=\columnwidth]{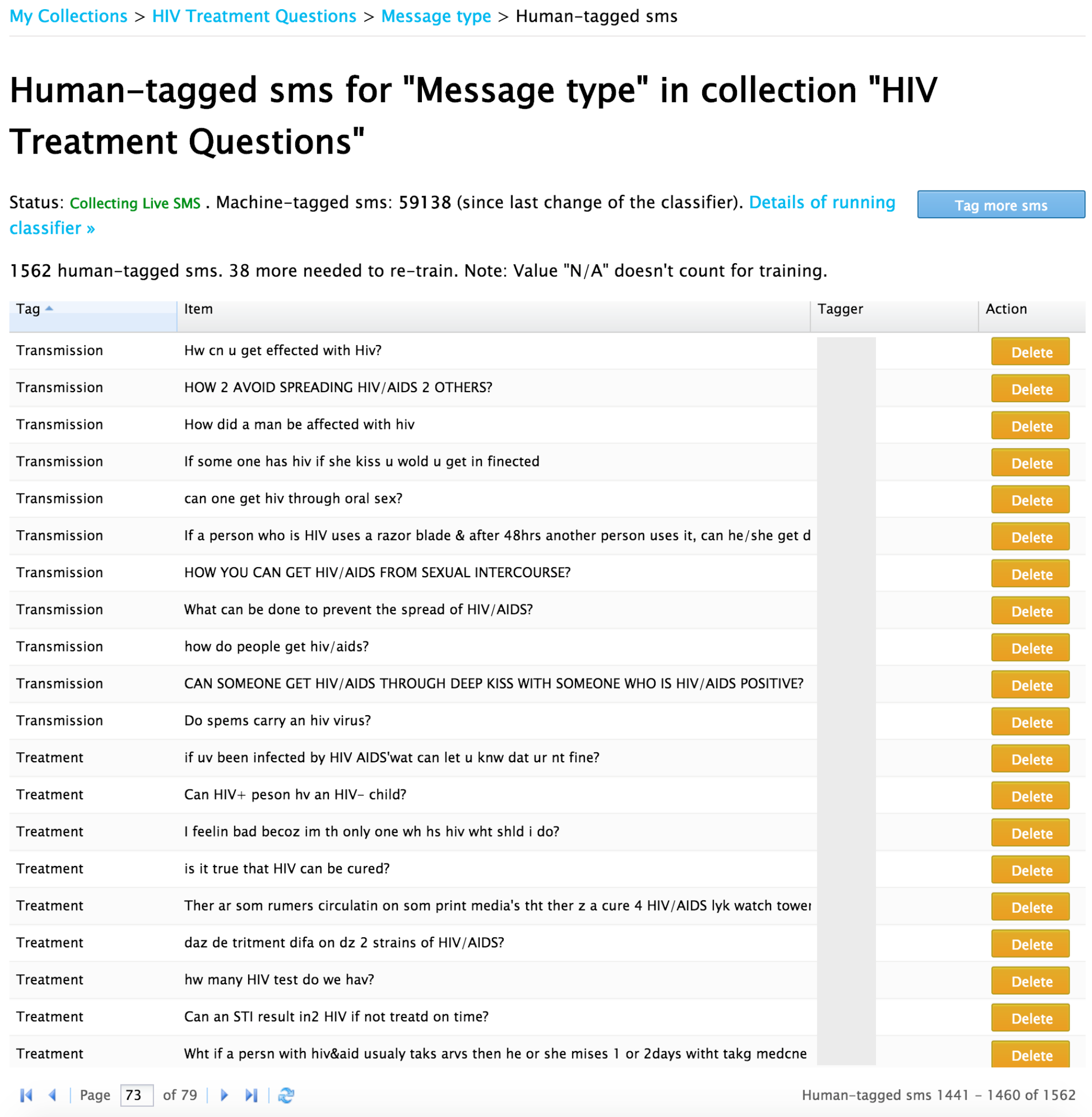}
    \caption{Human-tagged examples interface (human-tagger name is hidden for privacy purposes)\label{fig:labels}}
\end{figure}

Moreover, the task generator module implements a de-duplication strategy, which is an intermediate step to ensure that only novel tasks are selected for labeling. %Next, human labelers examine and provide an appropriate label (i.e. a category from the list of UNICEF categorise shown in section 2.1) to each message. 
Figure~\ref{fig:labels} shows a few human labeled messages, which is also an interface for administrators of the systems to examine the quality of human labels. A wrongly labeled message can also be deleted from the same interface to prevent it being used as a training example. 

The system trains new models after receiving every 50 labeled messages. However, to achieve and maintain high classification accuracy, we also employ an active learning approach. That is, if a model is already trained, then the task generator module picks and priorities the tasks for which the machine confidence is low (in this case we set it to $\leq$0.60). This approach helps classifiers generalize concepts quickly. 

The \textit{learner} module uses subset of the human-labeled messages as training set (80\%) and learns a model. The model evaluation is performed on a hold-out test set consisting of remaining 20\% of the human-labeled messages. Once a model is trained it then enables the \textit{classifier} module to start classifying the incoming messages. The human labeling process continues until the expert observes an acceptable classification accuracy. And, during the machine classification, more training examples can be provided, if a decrease in classification accuracy is observed. 

Finally, machine classified messages are ranked based on their categories and the machine confidence score assigned to them. UNICEF expert simply downloads the classified messages from each category and direct them to corresponding response experts.

\section{Pilot Deployment and \\ Evaluation}
\label{sec:study}
To determine the effectiveness of the system, we performed a pilot deployment in conjunction with UNICEF. First, we provide a hands-on training session to two experts from UNICEF in which SMS-based collection and classifier creation processes were highlighted. During the pilot deployment, the system was tested using a real SMS dataset consisting of 60,000 messages collected by UNICEF Zambia on various heath-related issues. 

We developed a streaming system to mimic the functionality of the UNICEF U-Report system. The streaming system provides a stream of SMS messages to the SMS pusher. The SMS pusher system pushes the received SMS to AIDR-SMS. A subset of messages are pushed to the system and labeled by the UNICEF experts to be served as training set for building classifiers created using the categories listed in section 2.1. The system trains new models upon receiving new training examples (current setting is after every 50 labels). Classification accuracy increases as more training examples are becoming available.

\begin{figure}
 \includegraphics[width=\columnwidth]{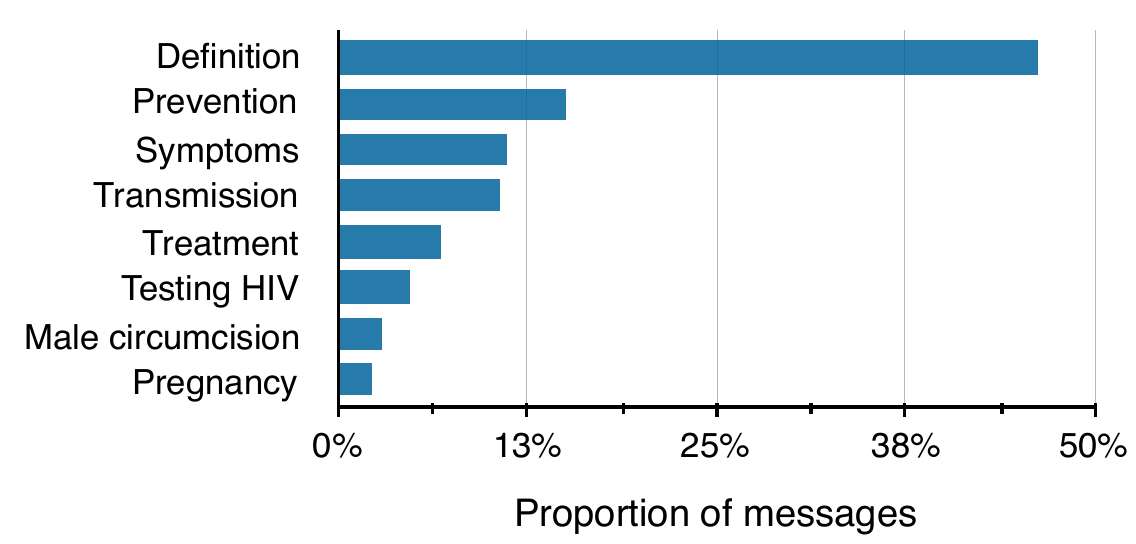}
   \caption{Proportion of SMS messages into different categories sorted in decending order\label{fig:proportion}}
\end{figure}

An acceptable accuracy level was obtained measured in terms of Area Under the ROC Curve\footnote{\url{https://en.wikipedia.org/wiki/Receiver_operating_characteristic}} (AUC = 82\%)\footnote{Higher numbers are better and 50\% represents a random classifier} after using a training set of 740 labeled messages. The system then classified the rest of the messages. Figure~\ref{fig:proportion} depicts the proportion of messages in different categories. As the deployment was managed by the UNICEF team itself, we asked them for a feedback about the system, which is given below.

\textbf{UNICEF feedback}: \textit{``Our overall experience with AIDR-SMS system was satisfactory. Our test found that the AIDR-SMS system offered high accuracy results on SMS that contained primarily a single topic or belonged to single category. This is to be expected and was explained to us by the AIDR-SMS team as being the result of the algorithm they use which can place the text in one or other category. The workflow we found that works well is then to ensure the messages needing categorization are simple and can contain primarily one topic..."}

\section{Related work}
\label{sec:relatedwork}
Retrospective approach for the analysis of continuous data streams is not a feasible solution. When rapid analysis of data is necessary, the data items must be processed as soon as they arrive. This requires real-time processing capabilities. Stream processing systems (e.g. \cite{neumeyer2010s4,abadi2003aurora}) perform computations on an unbounded, continuous, and time-varying data and if performed timely and effectively, can support real-time decision-making processes~\cite{carney2002monitoring}. However, a significant drawback of traditional stream processing systems is that they rely entirely on automated algorithms of data processing, and as such they are limited by the processing capabilities of these algorithms. Crowdsourcing, which is another core component of our approach, is a new computing paradigm that can be used when machines are not producing desirable results~\cite{yuen2011survey,geiger2011crowdsourcing}. Completely relying on crowd workers is also not an idea solution, as it is generally slower and more costly (i.e. in terms of time or money).

To overcome the above-mentioned issues, in this paper we introduce a hybrid system, which combines good characteristics of both types of systems (stream processing and crowdsourcing) to classify continuous data stream of short text messages while maintaining high classification accuracy. Moreover, research studies show that employing machine learning techniques to process short messages (e.g. SMS, tweets, etc.) produce better results than simply rule or keyword based approaches~\cite{imran2015processing}.

% !TEX root = paper.tex
\section{Conclusions \& Future Work}
\label{sec:conclusion}
In this paper, we presented a hybrid system that combines human and machine computation elements to classify stream of SMS messages into different categories. We employed supervised machine learning techniques to train multi-class classifiers using training data obtained from expert human workers. The system has been deployed and tested by the UNICEF team in conjunction with their U-Report platform. High classification accuracy and the positive feedback from UNICEF show the effectiveness of the system. 

However, the UNICEF team also suggested some improvements for the future development of the system. Two important suggestions as are follows: (i) add multi-label classification capabilities to the system in addition to its multi-class classification model, and (ii) allow batch training of a classifier that is useful in some scenarios (e.g. training semantically similar classifiers using past labels). In future, we plan to make the system more robust by implementing the suggested improvements.% baIn the future developments of the system, new features based on the UNICEF suggestions will be added to the system.

%\balance{} 

% \bibliographystyle{ACM-Reference-Format-Journals}
%\bibliographystyle{SIGCHI-Reference-Format}
\bibliographystyle{abbrv}
\bibliography{references}

\begin{thebibliography}{10}

\bibitem{abadi2003aurora}
D.~Abadi, D.~Carney, U.~Cetintemel, M.~Cherniack, C.~Convey, C.~Erwin,
  E.~Galvez, M.~Hatoun, A.~Maskey, A.~Rasin, et~al.
\newblock Aurora: a data stream management system.
\newblock In {\em Proc. of the 2003 ACM SIGMOD international conference on
  Management of data}, pages 666--666. ACM, 2003.

\bibitem{carney2002monitoring}
D.~Carney, U.~{\c{C}}etintemel, M.~Cherniack, C.~Convey, S.~Lee, G.~Seidman,
  M.~Stonebraker, N.~Tatbul, and S.~Zdonik.
\newblock Monitoring streams: a new class of data management applications.
\newblock In {\em Proc. of the 28th international conference on Very Large Data
  Bases}, pages 215--226. VLDB Endowment, 2002.

\bibitem{geiger2011crowdsourcing}
D.~Geiger, M.~Rosemann, and E.~Fielt.
\newblock Crowdsourcing information systems: a systems theory perspective.
\newblock In {\em Proc. of the 22nd Australasian Conference on Information
  Systems (ACIS 2011)}, 2011.

\bibitem{imran2015processing}
M.~Imran, C.~Castillo, F.~Diaz, and S.~Vieweg.
\newblock Processing social media messages in mass emergency: a survey.
\newblock {\em ACM Computing Surveys (CSUR)}, 47(4):67, 2015.

\bibitem{imran2014aidr}
M.~Imran, C.~Castillo, J.~Lucas, P.~Meier, and S.~Vieweg.
\newblock {AIDR}: Artificial intelligence for disaster response.
\newblock In {\em Proc. of the companion publication of the 23rd international
  conference on World wide web companion}, pages 159--162. International World
  Wide Web Conferences Steering Committee, 2014.

\bibitem{imran2014coordinating}
M.~Imran, C.~Castillo, J.~Lucas, M.~Patrick, and J.~Rogstadius.
\newblock Coordinating human and machine intelligence to classify microblog
  communications in crises.
\newblock {\em Proc. of ISCRAM}, 2014.

\bibitem{liaw2002classification}
A.~Liaw and M.~Wiener.
\newblock Classification and regression by randomforest.
\newblock {\em R news}, 2(3):18--22, 2002.

\bibitem{melville2013amplifying}
P.~Melville, V.~Chenthamarakshan, R.~D. Lawrence, J.~Powell, M.~Mugisha,
  S.~Sapra, R.~Anandan, and S.~Assefa.
\newblock Amplifying the voice of youth in {Africa} via text analytics.
\newblock In {\em Proc. of the 19th ACM SIGKDD international conference on
  Knowledge discovery and data mining}, pages 1204--1212. ACM, 2013.

\bibitem{neumeyer2010s4}
L.~Neumeyer, B.~Robbins, A.~Nair, and A.~Kesari.
\newblock S4: Distributed stream computing platform.
\newblock In {\em Data Mining Workshops (ICDMW), 2010 IEEE International
  Conference on}, pages 170--177. IEEE, 2010.

\bibitem{sriram2010short}
B.~Sriram, D.~Fuhry, E.~Demir, H.~Ferhatosmanoglu, and M.~Demirbas.
\newblock Short text classification in {Twitter} to improve information
  filtering.
\newblock In {\em Proc. of the 33rd international ACM SIGIR conference on
  Research and development in information retrieval}, pages 841--842. ACM,
  2010.

\bibitem{yuen2011survey}
M.-C. Yuen, I.~King, and K.-S. Leung.
\newblock A survey of crowdsourcing systems.
\newblock In {\em Privacy, Security, Risk and Trust (PASSAT) and 2011 IEEE
  Third Inernational Conference on Social Computing (SocialCom), 2011 IEEE
  Third International Conference on}, pages 766--773. IEEE, 2011.

\end{thebibliography}

\end{document}